\documentclass[epj]{svjour}
\usepackage{amsmath}
\usepackage{graphics}
\usepackage{dcolumn}
\usepackage{bm}

\newcommand{\lw}[1]{\smash{\lower 1.5ex\hbox{#1}}}
\begin{document}
\title{Modified Hagedorn formula including temperature fluctuation}
\subtitle{-Estimation of temperatures at RHIC experiments-}
\author{M.~Biyajima\inst{1}\thanks{\emph{e-mail:}
biyajima@azusa.shinshu-u.ac.jp} \and
T.~Mizoguchi\inst{2}\thanks{\emph{e-mail:} mizoguti@toba-cmt.ac.jp} \and
N. Nakajima\inst{3}\thanks{\emph{e-mail:} nakajimn@med.kochi-u.ac.jp}
\and N. Suzuki\inst{4}\thanks{\emph{e-mail:}suzuki@matsu.ac.jp} \and
G.~Wilk\inst{5}\thanks{\emph{e-mail:} wilk@fuw.edu.pl}
}                     
%
%
\institute{School of General Education, Shinshu University Matsumoto
390-8621, Japan \and Toba National College of Maritime Technology, Toba
517-8501, Japan \and Center of Medical Information Science, Kochi
University,
 Kochi, 783-8505, Japan \and
 Department of Comprehensive Management,
             Matsumoto University, Matsumoto 390-1295,
             Japan \and The Andrzej So\l tan Institute for Nuclear Studies,
Ho\.za 69, 00681 Warsaw, Poland}
\date{Received: date / Revised version: date}
%
\abstract{ We have systematically estimated the possible temperatures
obtained from an analysis of recent data on $p_t$ distributions observed
at RHIC experiments. Using the fact that observed $p_t$ distributions
cannot be described by the original Hagedorn formula in the whole range
of transverse momenta (in particular above $6$ GeV/c), we propose a
modified Hagedorn formula including temperature fluctuation. We show that
by using it we can fit $p_t$ distributions in the whole range and can
estimate consistently the relevant temperatures, including their
fluctuations.
} 
\maketitle

%
%

\section{Introduction} \label{intro}

One of the characteristic features in every high energy collision
experiment is the production of large numbers of secondaries (mostly
pions). From the very beginning of the history of the multiparticle
production processes, it was realized that a possible way to treat them
was to employ some sort of statistical approach \cite{STAT}. This idea
found its most mature formulation in the statistical bootstrap model
proposed by Hagedorn \cite{Hagedorn}, in which the exponential growth of
the number of hadronic resonances with mass is one of the most
fundamental issues~\cite{HuangNahmF}. The proposed formula is
\begin{eqnarray}
    \frac{d^3\sigma}{dp^3} = C \int dm \rho(m)
        \exp\left(-\sqrt{p_l^2 + p_t^2 + m^2} \beta_0\right)\,.  \label{eq1}
\end{eqnarray}
In Eq.(\ref{eq1}), $\rho(m)$ denotes the density of resonances given by
\begin{eqnarray}
\rho(m) = \frac{\exp(m\beta_H)}{(m^2 + m_0^2)^{5/4}},  \label{eq2}
\end{eqnarray}
where $\beta_H = 1/(k_BT_H)$, the inverse of the so called Hagedorn
temperature $T_H$, is a parameter to be deduced from data on resonance
production \cite{explanat}. The other parameter is $\beta_0 =
1/(k_BT_0)$, with $T_0$ explicitly governing the observed energy
distribution and therefore identified with the {\it temperature of the
hadronizing system}. In the followings we put $k_B=1$. One of the aims in
the study of multiparticle production processes is therefore the best
possible estimation of this quantity. To this end we would like to
investigate the measured transverse momentum ($p_t$) distributions
integrated over longitudinal degrees of freedom. From Eq.(\ref{eq1}), we
have
\begin{eqnarray}
 S_0 \equiv \frac{d^2\sigma}{2 \pi p_t dp_t}
      = C\int dm \rho(m) m_t K_1(m_t \beta_0)\,,   \label{eq3}
\end{eqnarray}
where $ m_t = \sqrt{p_t^2 + m^2} $ is transverse mass and $K_1$ is the
Bessel function.

However, as was recently demonstrated by us \cite{Biyajima:2004ub}, this
simple formula can explain the RHIC data only in the limited range of
transverse momenta, namely for $p_t \leq 6$ GeV/c. For larger values of
$p_t$ data exhibit a power-like tail. There are many attempts to explain
it using some kind of nonequilibrium approach like, for example, the flow
or decay of resonances (see \cite{FloRes} for most recent review and
further references); instead of trying to exclude them we would like to
investigate the possibility that the observed nonexponential spectra
could result from some form of equilibrium characteristic of nonextensive
thermodynamics. In fact, as was shown in \cite{Biyajima:2004ub}, using an
approach based either on nonextensive statistics or on stochastic
approach one can successfully account for the whole range of the observed
transverse momenta. The reason for this success is the fact that in both
approaches the resultant distributions are {\it intrinsically
non-exponential}, ranging from a power-law like form (cf. Eq.(\ref{eq5})
below) to a gaussian in transverse rapidity \cite{MC} (which can be
regarded as another implementation of the effective power-law
distribution)\footnote{ See also~\cite{suzu06}, where flow effect is
included and relation between Gaussian-like distribution in transverse
rapidity and power law behavior in $p_t$ is discussed.}. The fact that
the proposed formulas can fit {\it the whole range} of $p_t$ is by itself
very interesting and important observation as it shows that the power-law
is present not only in very hard scale physics but that it reflects also
a possible nontrivial property of hadronic matter in equilibrium (like,
for example, Quark Gluon Plasma) \cite{QM2005}.

Such properties are best seen in an approach using a nonextensive
statistical model in which two parameters are used: the action of the
heat bath is described now by the mean temperature $T_0$ and by the
nonextensivity parameter $q$, which can be identified with some specific
intrinsic fluctuations of the temperature existing in the hadronizing
system under consideration \cite{Wilk:1999dr}. In the case when these
fluctuations can be described by gamma distribution one can write exact
formulas \cite{Wilk:1999dr} telling us that (cf.~\cite{Wilk:1999dr})
\begin{eqnarray}
  [1 -(1-q)\beta_0 H_0]^{\frac 1{(1-q)}} &=& \int_0^{\infty}
  e^{-\beta H_0}f_{\Gamma}(\beta)d\beta\,, \label{eq4}
\end{eqnarray}
where
\begin{eqnarray}
  f_{\Gamma}(\beta) &=& \frac{1}{\Gamma(\alpha)}
   \Bigl(\frac{\alpha}{\beta_0}\Bigr)^{\alpha}
    \beta^{\alpha-1}e^{-\frac{\alpha}{\beta_0}\beta}\,, \quad \alpha
    =\frac{1}{q-1} \, .\label{eq4b}
\end{eqnarray}
In general, one refers to the concept of so called {\it superstatistics}
introduced in \cite{BC}. In our previous work~\cite{Biyajima:2004ub},
 RHIC data were described  by the following
distribution with $H_0 = \sqrt{p_l^2 +m_t^2}$:
\begin{eqnarray}
 \frac{d^2\sigma}{2 \pi p_t dp_t} = C\int_0^{\infty}dp_l\left[1-\frac{1-q}{T_0}
   \sqrt{p_l^2+m_t^2}\right]^{\frac{1}{1-q}}\,.   \label{eq5}
\end{eqnarray}
As is seen in~\cite{Biyajima:2004ub}, this formula leads to very good
agreement with all RHIC data~\cite{STAR,BRAHMS,PHENIX}.

It is important to notice that Eq.~(\ref{eq5}) has essentially the same
form as the formula proposed long time ago and used with success in many
QCD-inspired power-law fits to experimental data~\cite{CM,UA1,Hagpt}
(recently used also by RHIC collaborations \cite{Stillment}):
\begin{eqnarray}
  \left( 1 + \frac{p_t}{p_0}\right)^{-n} \longrightarrow \left\{
 \begin{array}{l}
  \exp\left(-\frac n{p_0}p_t\right)\quad {\rm for}\ p_t \to 0, \smallskip\\
  \left(\frac{p_0}{p_t}\right)^n\quad {\rm for}\ p_t \to \infty,
 \end{array}
 \right .  \label{eq6} .
\end{eqnarray}
However, one has also to realize the important difference in physical
pictures leading to Eq.~(\ref{eq5}) and Eq.~(\ref{eq6}). The underlying
physical picture in Eq.~(\ref{eq6}) is that the small $p_t$ region is
governed by {\it soft physics} described by some unknown unperturbative
theory or model, and the large $p_t$ region is governed by {\it hard
physics} believed to be described by perturbative QCD.
Contrary to it, the nonextensive formula Eq.(\ref{eq5}), which is valid
in the whole range of $p_t$, does not claim to originate from any
particular theory. It merely offers the kind of general unifying
principle, namely the existence of some kind of complicated equilibrium
involving all scales of $p_t$, which is described by two parameters,
$T_0$ and $q$: the temperature $T_0$ describing its mean properties and
the parameter $q$ describing action of the possible nontrivial long range
effects believed to be caused by fluctuations but essentially also by
some correlations or long memory effects \cite{T} \footnote{The origin of
such fluctuations and/or correlations must be most probably traced back
to the nonperturbative QCD, cf., for example, \cite{QCDfluct}.}.

\section{Calculations and results}\label{calculations}

In this paper, we would like to compare results of an analysis of $p_t$
spectra measured at RHIC experiments \cite{STAR,BRAHMS,PHENIX} performed
by using three approaches: the original Hagedorn model, Eq.~(\ref{eq3}),
the QCD-inspired power-like formula, Eq. (\ref{eq6}), and the modified
Hagedorn formula including temperature fluctuation given by :
\begin{eqnarray}
S_{tot} &\equiv& \frac{d^2\sigma}{2 \pi p_t dp_t} = C\! \int \!dy \cosh y
\int\! dm \rho (m)m_t \cdot \nonumber\\
&&\quad \qquad\qquad \cdot [1- \beta_0 (1-q) m_t \cosh
y]^{\frac{1}{1-q}}\,\,. \label{eq8}
\end{eqnarray}
It can be written also in the form of series ($\alpha = 1/(q-1)$):
\begin{eqnarray}
 S_{tot} &=& \frac{4C}{\alpha-1}\int_{m_\pi}^\infty dm \rho(m)
    \frac{\beta_0 m_t^2/\alpha}{(1+\beta_0
    m_t/\alpha)^\alpha}\cdot  \label{hage1}\\
    && \cdot \sum_{k=0}^{\infty} \frac{\Gamma(k+3/2)\Gamma(\alpha +1 + k)}
       {\Gamma(\alpha + k + 1/2)\Gamma(k+1)}
    \left(\frac{1-\beta_0 m_t/\alpha}{1+\beta_0 m_t/\alpha}\right)^k\,,
     \nonumber
\end{eqnarray}
or, accounting for the smallness of $q-1$ encountered in our fits
and of the fact that we are interested only in midrapidity region
(i.e., for small $y$) one can write it also as\footnote{In our case,
because we are integrating over the whole mass spectrum  $\rho(m)$
in Hagedorn formula, we cannot simply expand in $(q-1)$ and keep
only linear term as it is done on such occasions in the literature,
cf., for example \cite{S0}, because for large masses $m$ the series
becomes divergent.}
\begin{eqnarray}
S_{tot} &\simeq& C \int dy \cosh y \int dm \rho (m) m_t\cdot
\nonumber\\
&& \cdot \left[ 1 + \frac{1}{2}(q-1)\beta_0^2  m_t^2 y^2\right]\cdot\nonumber\\
&& \cdot \exp\left[-\beta_0 m_t \cosh y + \frac{1}{2}(q-1)\beta_0^2
m_t^2\right] \, . \label{hage1b}
\end{eqnarray}
Equations (\ref{hage1}) and (\ref{hage1b})  are used to check the
numerical integration of Eq.(\ref{eq8}).
\begin{figure*}
  \begin{center}
  \resizebox{0.41\textwidth}{!}{\includegraphics{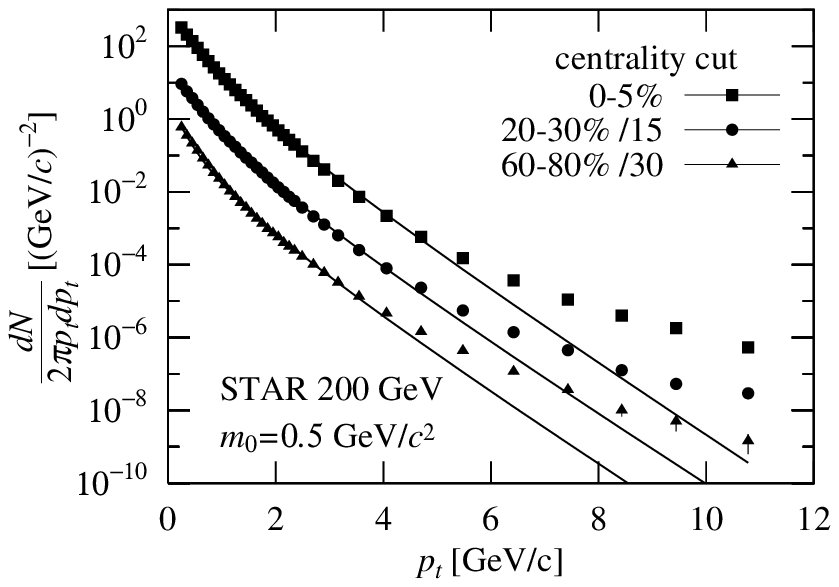}}
  \resizebox{0.41\textwidth}{!}{\includegraphics{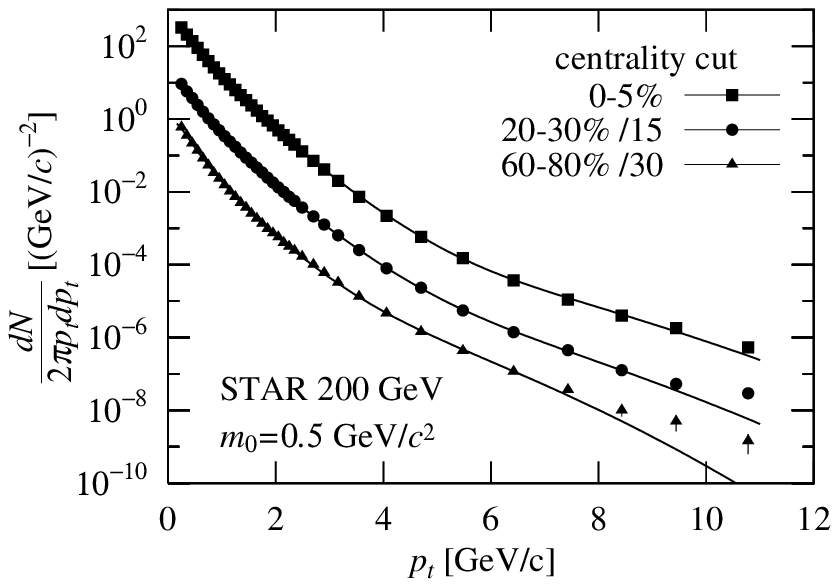}}
  \end{center}
  \caption{Analysis of STAR data~\cite{STAR} by using usual
 Hagedorn formula (Eq.(\ref{eq3}), left panel) and its nonextensive
 generalization (Eq.~(\ref{eq8}), right panel).}
  \label{Fig1}
\end{figure*}
\begin{table*}
 \caption{Parameters of our analysis presented in Fig.~\ref{Fig1} (left
 panel) by the use of Eq.~(\ref{eq3}) which corresponds to $q-1=0$ in Eq.(\ref{eq8}).
 Those for right panel with $q-1 \neq 1$ in Eq.(\ref{eq8}) can be found in Table 4.
 Other parameters are $m_0=0.5$ GeV (fixed), $\delta T_H=$ 0.0001-0.002 and $\delta T_0 =$
 0.0001-0.002. Notice that very large values of $\chi^2$are obtained for fits with
 $q-1=0$.}
 \begin{center}
  \renewcommand{\arraystretch}{0.8}
  \begin{tabular}{|c|cccc|}
   \hline
    C.\ C.\ (\%) & $C$ &  $T_H$ (GeV) & $T_0$ (GeV) & $\chi^2$/n.d.f.\\
   \hline & & & &  \\
    0-5   & 816$\pm$ 15   & 0.086 & 0.085 & 532/32 \\
    20-30 & 382$\pm$7    & 0.077 & 0.076 & 249/32 \\
    60-80 & 106$\pm$ 2  & 0.037 & 0.037 & 308/32 \\
    & & & &  \\
  \hline
 \end{tabular}
\end{center}
\label{Table1}
\end{table*}
\vspace{5mm}
\begin{figure*}
  \begin{center}
  \resizebox{0.41\textwidth}{!}{\includegraphics{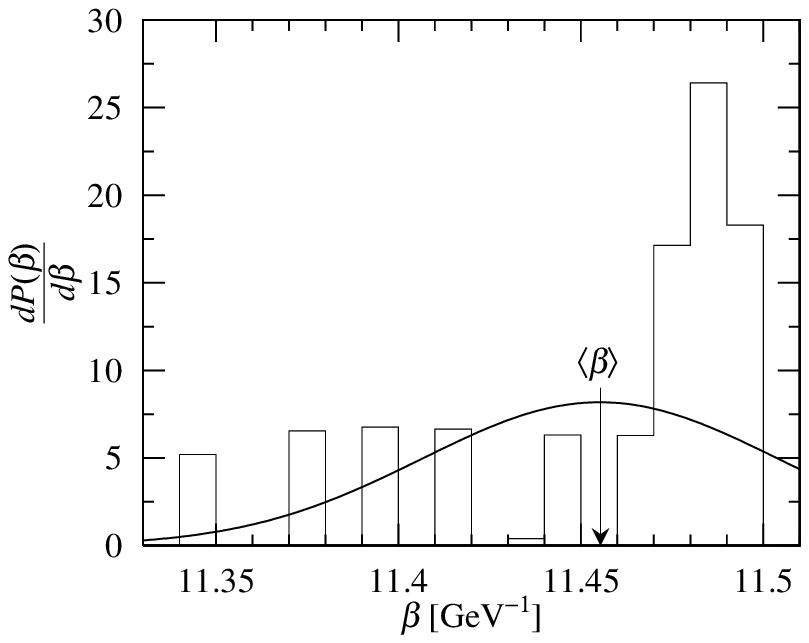}}
  \resizebox{0.43\textwidth}{!}{\includegraphics{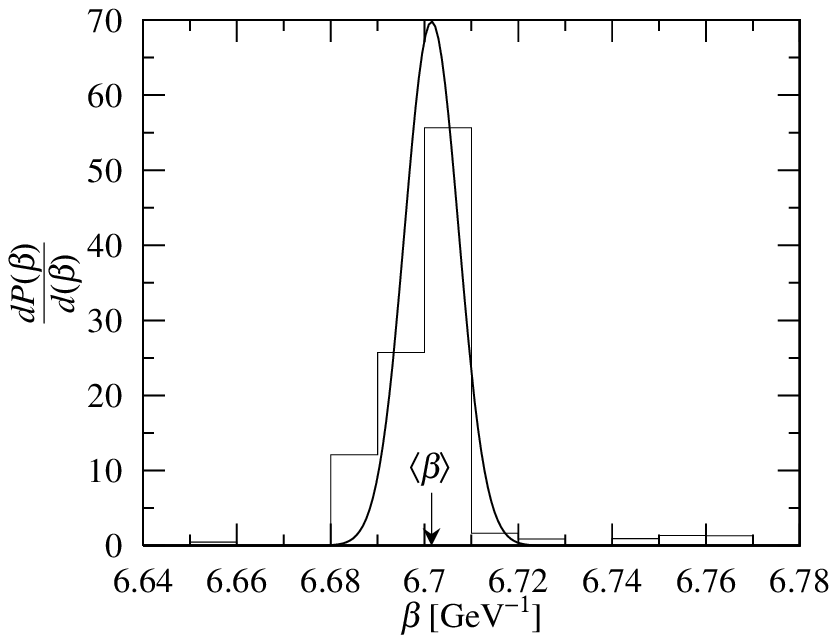}}
  \end{center}
  \caption{Temperature fluctuation in STAR data (for C.C.
$=0-5$ \%) \protect\cite{STAR} are analyzed by the use of Eq.~(\ref{eq3})
with $q-1=0$ (left panel) and Eq. (\ref{eq8}) with $q-1\neq 0$ (right
panel).}
  \label{Figure1a}
\end{figure*}

At first the STAR data~\cite{STAR} were analyzed using of Eq.~(\ref{eq3})
(which corresponds to $q=1$ in Eq.~(\ref{eq8})) and modified Hagedorn
formula, Eq.~(\ref{eq8}). Results are shown in Fig.~\ref{Fig1} and
Table~\ref{Table1}. The corresponding results for BRAHMS and PHENIX
data~\cite{BRAHMS,PHENIX} are very similar. As can be seen in
Fig.~\ref{Fig1}, whereas distributions in the small $p_t$ region can be
explained by simple formula (\ref{eq3}), data including the larger $p_t$
region can only be explained by using the modified Hagedorn formula,
Eq.~(\ref{eq8}) (or Eqs.~(\ref{hage1}) and (\ref{hage1b})). The nonzero
values of $|q-1|$ are then interpreted  as an indication of the sizeable
temperature fluctuations existing in the hadronizing system
\cite{Wilk:1999dr,BC}.

\begin{figure*}
  \begin{center}
  \resizebox{0.41\textwidth}{!}{\includegraphics{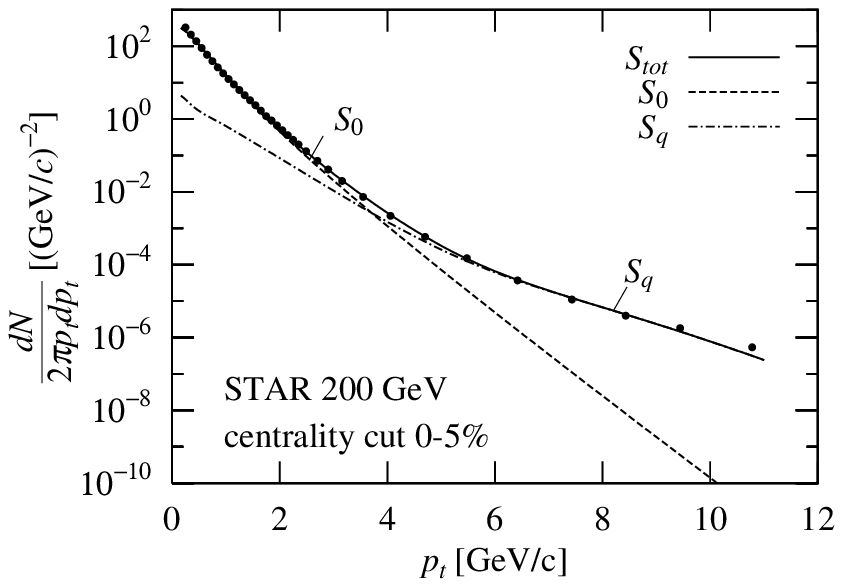}}
  \resizebox{0.41\textwidth}{!}{\includegraphics{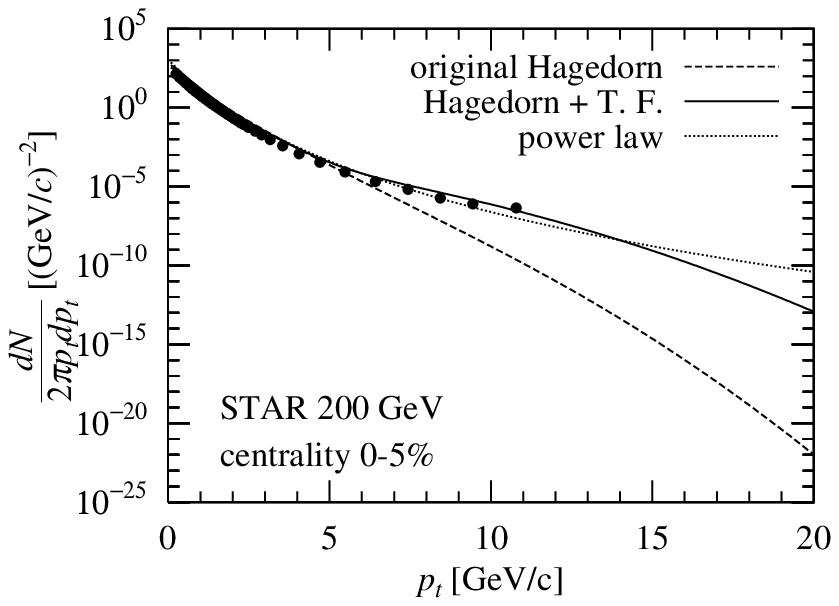}}
  \resizebox{0.41\textwidth}{!}{\includegraphics{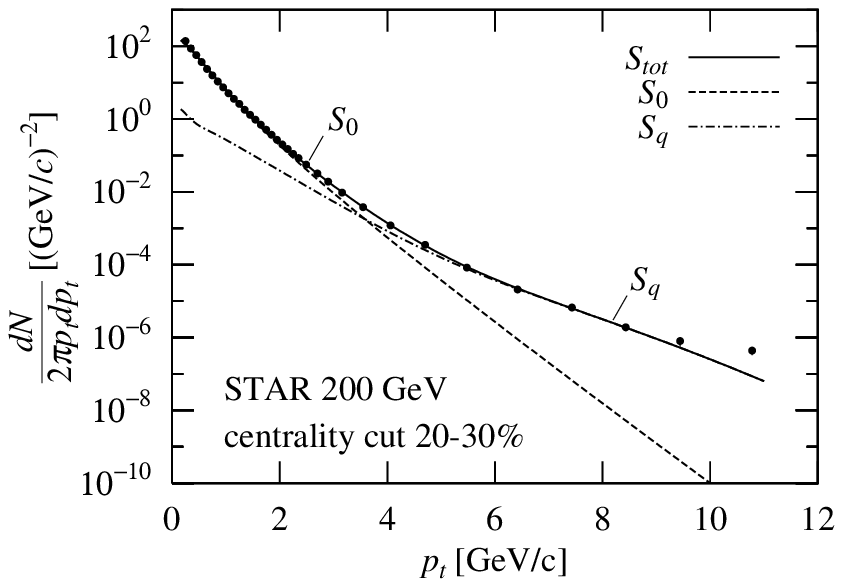}}
  \resizebox{0.41\textwidth}{!}{\includegraphics{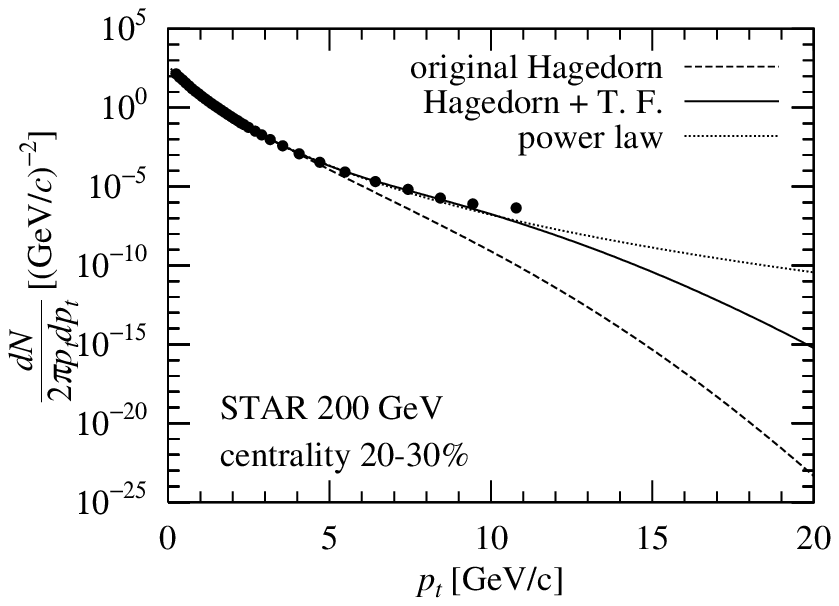}}
  \resizebox{0.41\textwidth}{!}{\includegraphics{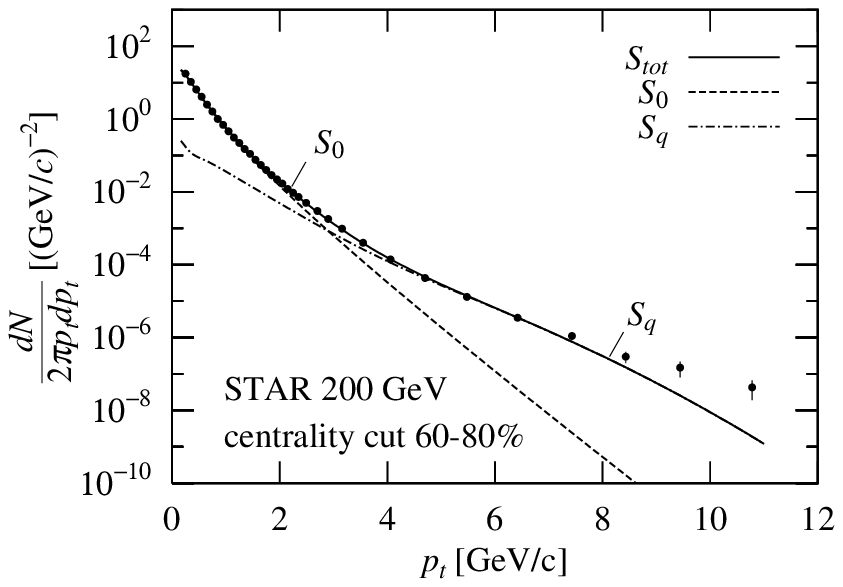}}
  \resizebox{0.41\textwidth}{!}{\includegraphics{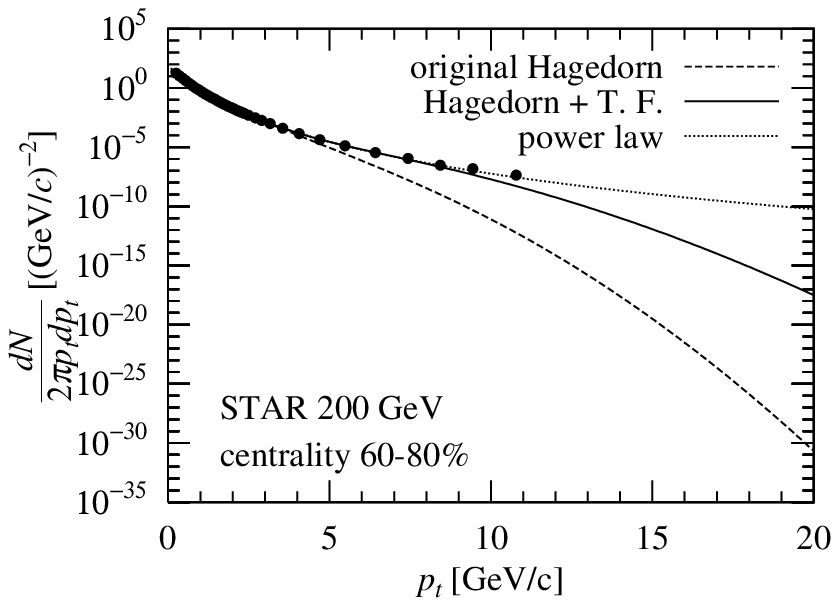}}
  \end{center}
  \caption{Example of the visualization of results presented
in Tables 2 and 4 using STAR \cite{STAR} results. Left panels show in
detail contribution of different mechanisms represented by $S_{tot}$ Eq.
(\ref{eq8}), by $S_0$ Eq. (\ref{eq3}) and by their difference denoted by
$S_q=S_{tot}-S_0$. Right panels show results on original Hagedorn model,
Eq.(\ref{eq1}) by dashed line, the modified the Hagedorn formula
including temperature fluctuation, Eq.(\ref{eq8}), by solid line, and
Eq.(\ref{eq6}) by dotted line.
           }
  \label{Fig2}
\end{figure*}

In fact RHIC data allow us to investigate the fluctuation of temperature
in more detail, cf. Fig. \ref{Figure1a}. The centrality cut C.C. $=0-5$
\% region of STAR data \cite{STAR} was fitted by using, respectively,
Eq.~(\ref{eq3}) (with $q-1=0$, left panel) and Eq.~(\ref{eq8}) (with $q$
as in Table 4, right panel). Fit was performed by fixing all parameters
in Eqs.~(\ref{eq3}) and (\ref{eq8}) except $\beta_0$; $\beta(=\beta_0)$
is then calculated for each of $35$ data points and it is assumed that
the reciprocal of the each error bar calculated by the fitting program
MINUIT is proportional to the corresponding probability of this value of
$\beta$, $P(\beta)$. In this way a probability distribution for $\beta$
is obtained and presented in the form of a histogram in
Fig.\ref{Figure1a}. The histogram in each panel is then fitted to the
Gamma distribution with $\alpha=55000$ (shown as solid curves). The mean
value $\langle \beta\rangle$ is also shown in each panel in
Fig.\ref{Figure1a}. As can be seen, a good fit can be obtained {\it only}
when Eq.~(\ref{eq8}) is used and in this case the resultant distribution
of temperatures is very narrow. This result suggests that accounting for
intrinsic fluctuations considerably narrows the distribution of
temperatures (actually its reverse, $\beta = 1/T$) and minimizes what can
be regarded as a kind of systematic error in deduction of $\beta_0$ from
experimental data. Therefore it strongly suggests that the modified
Hagedorn formula, Eq.~(\ref{eq8}), should be used whenever possible.

The results of our fits to RHIC data~\cite{STAR,BRAHMS,PHENIX}
performed by using Eq.~(\ref{eq5}) (as given by nonextensive
statistical approach), Eq.~(\ref{eq6}) (representing the
QCD-inspired power-law formula) and Eq.~(\ref{eq8}) (given by the
modified Hagedorn formula proposed by us here) are presented in,
respectively, Tables 2, 3 and 4. The results for STAR data are then
also shown in Fig.\ref{Fig2}. In particular, the left hand panels of
Fig.~\ref{Fig2} demonstrate contribution of different mechanism
represented, respectively by $S_{tot}$ and $S_0$.  It is clear that
data for the larger $p_t$ region can be explained only by $S_{tot}$,
which can be attributed to the intrinsic primordial temperature
fluctuations in the hadronizing system. However, at present it is
difficult to treat this as a possible signal of a Quark-Gluon
Plasma. Notice that the temperature parameter $T_0=1/\beta_0$ in
Table 2 and 4 was estimated by the use of Eq.~(\ref{eq8}) from the
whole region of transverse momenta, whereas $\tilde{T}_0=p_0/n$,
which corresponds to temperature in Eq.~(\ref{eq6}), shown in Table
3 governs only the small $p_t$ region. RHIC data show that we always
have $\tilde{T}_0 > T_0$, i.e., that inclusion of fluctuations and
long-range correlations present in the hadronizing system lowers the
estimated value of its mean temperature. From Table 4, we can see
that both temperatures, $T_H$ and $T_0$, estimated by the use of
Eq.~(\ref{eq8}) decrease as the centrality cut, C.C., increases
(i.e., it can be argued that they increase with the volume of
interaction; similar effect concerning $T_H$ has been also found in
\cite{BP}). It should be emphasized that when one uses the modified
Hagedorn formula, Eq.~(\ref{eq8}), then  $T_H \sim T_0 \sim
m_{\pi}$, i.e., estimated values of $T_H$ and $T_0$ are almost equal
to $m_{\pi}$, which we regard as very reasonably
result\footnote{Actually, analysis performed assuming both thermal
and chemical equilibrium and including also baryons performed by GSI
group \cite{BM} gives $T=170$ MeV. In our case we are considering
only pions and get $T\simeq m_{\pi}$. This difference is important
for the description of phase diagram and we plan to address it
elsewhere. One more remark is in order here. The $T_0$ parameters
obtained by us are in a range of $T_c = 170$ MeV, the QCD crossover
temperature. On the other hand, traditional exponential fits for the
low $p_T$ part of pion spectra used to give $T = 340$ MeV, pointing
to a transverse flow with a Doppler blue-shift factor of two.
However, we do not claim that there is no transverse flow in RHIC
experiments, we only show that nonextensive approach can mimic this
effect as well.}.

\begin{table*}[h]
\caption{Analysis of RHIC data \cite{STAR,BRAHMS,PHENIX} by means of
nonextensive approach as given by Eq. (\ref{eq5}). For comparison the
results for $pp$ collisions are also shown.}
\begin{center}
\begin{tabular}{rccccc} \hline\hline
Coll. \& C.C.  &  $c$          & $T_0$ [GeV]      & $q$     & $1/(q-1)$ & $\chi^2/$n.d.f \\
\hline
BRAHMS 0-10\%  & 936$\pm$68  & 0.227$\pm$0.005 & 1.0394$\pm$0.0026 & 25.4 & 10.2/23 \\
       10-20\% & 716$\pm$56  & 0.217$\pm$0.005 & 1.0455$\pm$0.0029 & 22.0 & 12.9/23 \\
       20-40\% & 468$\pm$41  & 0.208$\pm$0.006 & 1.0507$\pm$0.0033 & 19.7 & 12.8/23 \\
       40-60\% & 265$\pm$32  & 0.185$\pm$0.007 & 1.0607$\pm$0.0044 & 16.5 & 10.6/23 \\
       60-80\% &36.2$\pm$4.2 & 0.165$\pm$0.005 & 1.0764$\pm$0.0024 & 13.1 & 2.76/23 \\
\hline
PHENIX\quad 0-5\% & 1530$\pm$359  & 0.195$\pm$0.012 & 1.0461$\pm$0.0060 & 21.7 & 5.00/29 \\
          5-15\%  & 1200$\pm$276  & 0.193$\pm$0.012 & 1.0472$\pm$0.0057 & 21.2 & 3.56/29 \\
          15-30\% &  760$\pm$180  & 0.189$\pm$0.012 & 1.0503$\pm$0.0058 & 19.9 & 5.50/29 \\
          30-60\%  &  384$\pm$96   & 0.170$\pm$0.011 & 1.0613$\pm$0.0055 & 16.3 & 2.60/29 \\
          60-80\% &  120$\pm$39   & 0.144$\pm$0.012 & 1.0728$\pm$0.0067 & 13.7 & 10.5/29 \\
          80-92\% & 59.2$\pm$32.0 & 0.114$\pm$0.017 & 1.0879$\pm$0.0106 & 11.4 & 8.99/29 \\
\hline
STAR\qquad 0-5\% & 3980$\pm$186 & 0.164$\pm$0.002 & 1.0651$\pm$0.0009 & 15.4 &  172/32 \\
      5-10\% & 2900$\pm$148 & 0.169$\pm$0.002 & 1.0622$\pm$0.0011 & 16.1 & 64.5/32 \\
     10-20\% & 2340$\pm$114 & 0.164$\pm$0.002 & 1.0662$\pm$0.0011 & 15.1 & 66.4/32 \\
     20-30\% & 1630$\pm$81  & 0.162$\pm$0.002 & 1.0684$\pm$0.0011 & 14.6 & 40.7/32 \\
     30-40\% & 1170$\pm$61  & 0.158$\pm$0.002 & 1.0709$\pm$0.0011 & 14.1 & 38.9/32 \\
     40-60\% &  739$\pm$39  & 0.146$\pm$0.002 & 1.0772$\pm$0.0010 & 13.0 & 14.7/32 \\
     60-80\% &  328$\pm$19  & 0.130$\pm$0.002 & 1.0850$\pm$0.0011 & 11.8 & 9.39/32 \\
    pp (nsd) & 49.9$\pm$5.5 & 0.111$\pm$0.003 & 1.0894$\pm$0.0014 & 11.2 & 10.1/29 \\
\hline
\end{tabular}
\end{center}
\label{t2}
\end{table*}
\vspace{-3mm}
\begin{table*}[h]
\begin{center}
\caption{Analysis of RHIC data \cite{STAR,BRAHMS,PHENIX} by means of the
QC-inspired power-like formula (\ref{eq6}). For comparison the results
for $pp$ collisions are also shown.}
\medskip
\begin{tabular}{rccccc} \hline\hline
Coll. \& C.C.  & $c$          & $n$          & $p_0$ [GeV]   &
                      $\tilde T = p_0/n$ & $\chi^2/$n.d.f \\
\hline
BRAHMS 0-10\%  &  353$\pm$19  & 32.2$\pm$3.3 & 8.89$\pm$1.09 & 0.276 & 7.96/23 \\
       10-20\% &  260$\pm$15  & 26.4$\pm$2.5 & 7.05$\pm$0.83 & 0.267 & 14.7/23 \\
       20-40\% &  163$\pm$11  & 22.8$\pm$2.1 & 5.87$\pm$0.70 & 0.257 & 13.9/23 \\
       40-60\% & 83.7$\pm$7.5 & 17.9$\pm$1.8 & 4.13$\pm$0.56 & 0.231 & 11.7/23 \\
       60-80\% & 11.1$\pm$1.0 & 12.8$\pm$0.5 & 2.58$\pm$0.17 & 0.202 & 2.86/23 \\
\hline
PHENIX\quad 0-5\% &  536$\pm$398  & 23.8$\pm$34.0 & 5.54$\pm$31.26 & 0.233 & 4.69/29 \\
          5-15\%  &  417$\pm$276  & 23.0$\pm$23.7 & 5.30$\pm$21.15 & 0.231 & 3.58/29 \\
          15-30\% &  260$\pm$149  & 21.3$\pm$13.2 & 4.84$\pm$11.01 & 0.227 & 5.54/29 \\
          30-60\%  &  120$\pm$924  & 16.6$\pm$91.8 & 3.40$\pm$90.58 & 0.205 & 2.66/29 \\
          60-80\% & 32.1$\pm$38.2 & 13.5$\pm$31.5 & 2.38$\pm$30.46 & 0.177 & 10.3/29 \\
          80-92\% & 12.8$\pm$35.4 & 10.8$\pm$31.6 & 1.53$\pm$30.44 & 0.142 & 8.83/29 \\
\hline
STAR\qquad 0-5\% & 1140$\pm$41  & 15.4$\pm$0.3 & 3.10$\pm$0.09 & 0.201 & 194/32  \\
          5-10\% &  843$\pm$33  & 16.5$\pm$0.4 & 3.44$\pm$0.12 & 0.208 & 68.6/32 \\
         10-20\% &  660$\pm$25  & 15.3$\pm$0.3 & 3.12$\pm$0.10 & 0.203 & 72.3/32 \\
         20-30\% &  457$\pm$18  & 14.7$\pm$0.3 & 2.94$\pm$0.09 & 0.200 & 42.8/32 \\
         30-40\% &  319$\pm$13  & 14.1$\pm$0.3 & 2.77$\pm$0.09 & 0.196 & 38.1/32 \\
         40-60\% &  190$\pm$8   & 12.6$\pm$0.2 & 2.30$\pm$0.07 & 0.182 & 13.9/32 \\
         60-80\% & 75.4$\pm$3.3 & 11.3$\pm$0.2 & 1.84$\pm$0.06 & 0.163 & 7.18/32 \\
        pp (nsd) & 10.8$\pm$0.9 & 10.4$\pm$0.2 & 1.42$\pm$0.06 & 0.136 & 11.6/29 \\
\hline
\end{tabular}
\end{center}
\label{t3}
\end{table*}

\begin{table*}
\begin{center}
\caption{Analysis of RHIC data \cite{STAR,BRAHMS,PHENIX} by means of
nonextensive modification of the Hagedorn formula as given by
Eq.~(\ref{eq8}). Maximum $m$ is fixed at 70 GeV (therefore in
(\ref{hage1}) one always has $(1- \beta_0m_t/\alpha)>0$). Numbers of
divisions for $y$ and $m$ in computations are given in the last column.
For comparison the results for $pp$ collisions are also shown.}
\begin{tabular}{rcccccc} \hline\hline
Coll. \& C.C. & $c$ & $q-1$ & $T_{\tiny H}$ [GeV] & $T_0$ [GeV] & $\chi^2/$n.d.f. & no. of div.\\
\hline BRAHMS 0-10\% &
156$\pm$3 & 0.00 & 0.192$\pm$0.000 & 0.178$\pm$0.000 & 15.4/22 & 6$\times$6\\

10-20\% & 106$\pm$5 & (4.76$\pm$0.62)$\times 10^{-4}$ &
0.206$\pm$0.007 & 0.187$\pm$0.005 & 13.1/22 &
6$\times$7\\

20-40\%  & 67.7$\pm$4.9  & (8.49$\pm$21.05)$\times 10^{-5}$ &
0.177$\pm$0.013 & 0.166$\pm$0.010 & 11.6/22 &
5$\times$3\\

40-60\% & 32.5$\pm$2.9 & (2.57$\pm$0.63)$\times 10^{-4}$ &
0.168$\pm$0.010 & 0.157$\pm$0.008 & 9.54/22 &
5$\times$4\\

60-80\% & 5.00$\pm$0.14 & (8.12$\pm$0.45)$\times 10^{-5}$ &
0.124$\pm$0.000 & 0.120$\pm$0.000 & 3.19/22 &
6$\times$6\\

\hline PHENIX\quad 0-5\% & 226$\pm$56 & (1.21$\pm$2.31)$\times
10^{-4}$ & 0.16$\pm$0.02 & 0.152$\pm$0.019 & 4.98/29 &
5$\times$6\\

5-15\% & 157$\pm$34 & (4.01$\pm$0.02)$\times 10^{-4}$ &
0.183$\pm$0.023 & 0.167$\pm$0.017 & 3.32/29 &
6$\times$5\\

15-30\% & 87.5$\pm$10.4 & (4.26$\pm$0.80)$\times 10^{-4}$ &
0.187$\pm$0.010 & 0.170$\pm$0.008 & 4.31/29 &
5$\times$3\\

30-60\% & 50.3$\pm$8.7 & (1.64$\pm$0.47)$\times 10^{-4}$ &
0.140$\pm$0.012 & 0.133$\pm$0.010 & 2.54/29 &
6$\times$7\\

60-80\% & 27.8$\pm$2.5 & (1.99$\pm$0.43)$\times 10^{-5}$ &
0.0731$\pm$0.0002 & 0.0719$\pm$0.0002 & 9.91/29 &
5$\times$6\\

80-92\% & 10.0$\pm$1.2 & (1.24$\pm$0.30$\times 10^{-5}$ &
0.0565$\pm$0.0002 & 0.0558$\pm$0.0001 & 8.71/29 &
12$\times$12\\

\hline STAR\qquad 0-5\% & 477$\pm$13 &     (1.48$\pm$0.05)$\times
10^{-4}$ & 0.140$\pm$0.001 & 0.132$\pm$0.001 & 56.6/31 &
6$\times$6\\

5-10\% & 443$\pm$15 & (1.08$\pm$0.06)$\times 10^{-4}$ &
0.127$\pm$0.002 & 0.122$\pm$0.002 & 38.0/31 &
7$\times$6\\

10-20\% & 326$\pm$17 & (1.02$\pm$0.10)$\times 10^{-4}$ &
0.126$\pm$0.004 & 0.121$\pm$0.003 & 33.8/31 &
6$\times$5\\

20-30\% & 236$\pm$14 & (8.15$\pm$1.00)$\times 10^{-5}$ &
0.119$\pm$0.004 & 0.115$\pm$0.004 & 30.0/31 &
6$\times$6\\

30-40\% & 169$\pm$10 & (7.13$\pm$0.09)$\times 10^{-4}$ &
0.113$\pm$0.004 & 0.109$\pm$0.004 & 25.5/31 &
6$\times$6\\

40-60\% & 109$\pm$4 & (4.40$\pm$0.22)$\times 10^{-5}$ &
0.0961$\pm$0.0014 & 0.0937$\pm$0.0013 & 24.5/31 &
6$\times$7\\

60-80\% & 46.0$\pm$1.0 & (2.80$\pm$0.08)$\times 10^{-5}$ &
0.0797$\pm$0.0001 & 0.0782$\pm$0.0000 & 23.4/31 &
6$\times$7\\

pp (nsd) & 4.98$\pm$0.15 & (2.87$\pm$0.08)$\times 10^{-5}$ &
0.0725$\pm$0.0000 & 0.0711$\pm$0.0000 & 38.2/28 &
20$\times$22\\

\hline
\end{tabular}
\end{center}
\label{t4}
\end{table*}

\section{Summary}\label{Summary}

We have presented a systematic analysis of RHIC data
\cite{STAR,BRAHMS,PHENIX} on transverse momenta distributions, which
allow, in principle, the deduction of the parameter believed to
represent the temperature $T_0$ of the hadronizing system. We have
shown that in order to fit the whole range of $p_t$ one has to use a
nonextensive approach, which accounts for temperature fluctuations
present in the hadronizing system. This has been compared with
approach using the old QCD-inspired power-like formulas introduced
long time ago. We have demonstrated that gradual accounting for the
intrinsic dynamical fluctuations in the hadronizing system by
switching from Eq.~(\ref{eq5}) (as given by nonextensive statistical
approach) to the modified Hagedorn formula including temperature
fluctuation, Eq.~(\ref{eq8}), substantially lowers the values of
parameter $q-1$. This is because part of the fluctuations ascribed
in Eq. (\ref{eq5}) to $q$ are accounted for by the resonance
spectrum $\rho(m)$ present in the Hagedorn formula. It also changes
the temperature we are looking for. Therefore one has to be very
careful when interpreting the temperature parameter obtained in such
fits, especially when attempting to address any questions concerning
Quark Gluon Plasma production issues\footnote{One should be aware of
the fact that there is still an ongoing discussion on the meaning of
the temperature in nonextensive systems. However, the small values
of the parameter $q$ deduced from data allow us to argue that, to
first approximation, $T_0$ can be regarded as the hadronizing
temperature in such system. One must only remember that in general
what we study here is not so much the state of equilibrium but
rather some kind of stationary state. For a thorough discussion of
the temperature of nonextensive systems, see \cite{Abe}. It is also
worth to be aware that in addition to the possibility of long-range
correlations and memory effects to be at work in relativistic
heavy-ion reactions (which were so far not yet proven) one can also
view $q > 1$ as a general, leading order finite-size effect, $q = 1
+ O(1/N)$ as proposed in \cite{Parvan}.}. If data with larger $p_t$
are available, we can further investigate whether the modified
Hagedorn formula including temperature fluctuation is really
applicable or not.

%
%
\section*{Acknowledgements}
This study is partially supported by Faculty of Science at Shinshu
University. One of authors (MB) would like to thank for fruitful
conversations at RITP in Kyoto University and RCNP in Osaka University.
Partial support of the Polish State Committee for Scientific Research
(KBN) (grant Nr 621/E-78/SPB/CERN/P-03/DWM 52/2004-2006 (GW)) is also
acknowledged.


\end{document}